\documentclass[12pt]{article}
\usepackage{ascmac}
\usepackage{ulem}
\usepackage{latexsym}
\usepackage[dvipdfmx]{graphicx}
\usepackage{amsmath}
\usepackage[top=3.5cm,bottom=3.5cm,left=2cm,right=2cm]{geometry}

\newcommand{\h}{\hspace}
\newcommand{\be}{\begin{equation}}
\newcommand{\e}{\end{equation}}
\begin{document}

\title{
\vbox{
\baselineskip 14pt
\hfill \hbox{\normalsize KUNS-2498}
} \vskip 1cm
\bf \Large Reconsideration of the Coleman's Baby Universe\vskip 0.5cm
}
\author{ Kiyoharu~Kawana\thanks{E-mail: \tt kiyokawa@gauge.scphys.kyoto-u.ac.jp}
\bigskip\\
\it \normalsize
 Department of Physics, Kyoto University, Kyoto 606-8502, Japan\\
\smallskip
}
\date{\today}

\maketitle

\abstract{\noindent\normalsize
}We reconsider the Coleman's mechanism which is a candidate of the solution to the Cosmological Constant Problem. We use the Lorentzian path integral and allow each universe has a different effective field theory and a vacuum. By using the probability distribution of coupling constants, it is shown that the cosmological constant of our universe does not necessary become small due to the effects of other universes.

\newpage

\section{Introduction}
Recently, there has been a development \cite{Kawai:2011rj,Kawai:2011qb,Kawai:2013wwa,Hamada:2014ofa,Hamada:2014xra} about the Cosmological Constant Problem (CCP) by the context of the Coleman's baby universe. In \cite{Kawai:2011rj,Kawai:2011qb,Kawai:2013wwa,Hamada:2014ofa,Hamada:2014xra}, the Lorentzian path integral has been used, and they have assumed that universes connected each other through the wormhole configurations have the same effective theory. In general, this effective theory can be different from the Standard Model (SM). They have also defined the wave function of multiverse, and found that our universe's density matrix is given by 
\be \rho(a^{'},a)=\int d\overrightarrow{\lambda}\omega(\overrightarrow{\lambda})^{2}|\mu|^{2}\phi_{0}^{*}(a^{'})\phi_{0}(a)\times\exp\left(\int da^{''}|\mu\phi_{0}(a^{''})|^{2}\right),\e
where $a$ is the radius of the universe, $\phi_{0}$ is the Wheeler-Dewitt wave function and $\mu$ is the probability amplitude of an universe emerging from nothing. In this set up, they have concluded that the coupling constants of our universe should be fixed at the point in the $\{\lambda_{i}\}$ space such that the integrand of Eq.(1) becomes maximum. Especially, it was shown that the Cosmological Constant (CC) should be zero asymptotically by this requiring. The important fact is that this conclusion is rather general in that it does not depend on the effective theory one assumes. 
Furthermore, it is also argued that other fine tuning problems\footnote{From the observation of the Higgs mass \cite{Chatrchyan:2012ufa,Aad:2012tfa}, it is argued that there might be no new physics between the weak and the Planck scale \cite{Hamada:2012bp,Hamada:2013cta,Hamada:2013mya,Hamada:2014iga,Hamada:2015ria}.} might be solved by the same mechanism\footnote{Of course, there are many other possibilities such as the multiple point principle \cite{Froggatt:1995rt}, the classical conformality \cite{Meissner:2006zh} and the asymptotic safety \cite{Shaposhnikov:2009pv} and so on.}. In this paper, fascinated by these works, we are trying to generalize their formalism in that we consider the possibility such that each universe has a different effective theory. 
In the following argument, we will show that, although it is actually possible to define the path integral of multiverse, 
the CC of our universe $\Lambda_{SM}$ does not necessary become small due to the effects of other universes. This conclusion is an opposition to the Coleman's first idea \cite{Coleman:1988tj} and the above recent works. Making more quantitative argument is very difficult because we need to know the histories of other universes. In this paper, we assume that macroscopic universes do not depend strongly on the microscopic theory describing the emergence of universes. As a result, such microscope effects can be represented by the probability amplitude $p_{\alpha}(j,K)$ with which a single universe having an effective theory $\alpha$, a topology $K$ and a vacuum $j$ emerges. The similar approaches are recently discussed in the context of the eternal inflation. See \cite{Garriga:2012bc,Vilenkin:2013ik,Vilenkin:2013loa} and the references there in. \\

This paper is organized as follows. In Section2, we define the multiverse path integral. In Section3, we construct the wave function of multiverse, and get the probability distribution as a function of the effective coupling constants by tracing out the number of universes and their scale factors. In Section4, we study a few examples and see that $\Lambda_{SM}$ is not necessary fixed zero due to the effects of other universes. 

\section{Path Integral of Multiverse}
\begin{figure}
\begin{center}
\includegraphics[width=10cm]{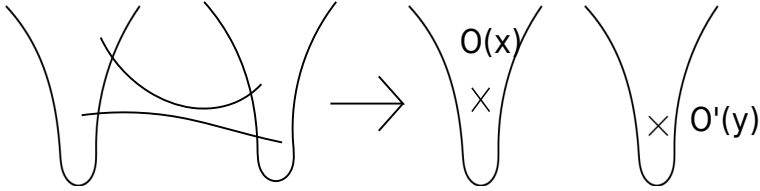}
\caption{An example of a wormhole configuration. Because the typical size of a wormhole is the Planck scale $M_{pl}$, the boundary of the wormhole can be understood as a local operator from the observers living in a universe.}
\label{mul1}
\end{center}
\end{figure}
In this section, we define the path integral of the multiverse where each universe can have a different effective theory. In the usual quantum filed theory, the path integral is defined by
\begin{equation}\sum_{all{\cal{M}}}\int_{{\cal{M}}} {\cal{D}}g{\cal{D}}\phi \hspace{2mm}\exp(-S^{G}_{E}-S^{M}_{E}),
\end{equation}
where $S^{G}_{E}$ is the Einstein Hilbert action, $S^{M}_{E}$ is the general action of the matter fields $\{\phi\}$ and ${\cal{M}}$ is the space-time manifold. The classical solution of the Euclidian gravity is called wormhole. We make the saddle point approximation of the wormhole configurations; the above fields can be divided to the classical wormhole configurations and the fluctuations around them:
\be g_{\mu\nu}(x)=g_{\mu\nu}^{(cl)}(x)+\tilde{g}_{\mu\nu}(x)\h{3mm},\h{3mm}\phi(x)=\phi_{cl}(x)+\tilde{\phi}(x).\label{eq:fluctuation}\e
The important fact is that the fields that make wormholes are common between the universes connected each other. In this paper, we assume that all the wormholes are produced by the common fields $\{\phi^{(c)}\}$. Let us focus on these common fields first. The fluctuations $(\tilde{g},\tilde{\phi})$ are divided into two types:\\
\\
(I)Fluctuations on the wormhole space time ${\cal{M}}_{worm}$:\\
\be (g_{\mu\nu}^{(W)}(x),\phi^{(W)}(x))\h{2mm},\h{2mm}x\in{\cal{M}}_{worm}.\e
(II)Fluctuations on large universes:\\
\be(g_{\mu\nu}^{(c)}(x),\phi^{(c)}(x))\h{2mm},\h{2mm}x\in{\cal{M}}_{n}:={\cal{M}}-{\cal{M}}_{worm},\e
where $n$ is the number of the universes.  When we integrate out the wormhole fluctuations, we must care that the boundary fields of these fluctuations are nothing but the fluctuations of the $n$-universes;
\be (g_{\mu\nu}^{(W)}(x),\phi^{(W)}(x))|_{{\cal{M}}_{n}\cap{\cal{M}}_{worm}}=(g_{\mu\nu}^{(c)}(x),\phi^{(c)}(x)).\e
For example, if we integrate the fluctuations of the wormhole having two legs (see Fig.\ref{mul1}), we obtain the factor
\be e^{-2S^{(W)}_{E}}c_{ij}\int_{{\cal{M}}_{n}}dx^{4}\int_{{\cal{M}}_{n}}dy^{4}\sqrt{g(x)}\sqrt{g(y)}{\cal{O}}^{i}(g^{(c)}(x),\phi^{(c)}(x))\times{\cal{O}}^{j}(g^{(c)}(y),\phi^{(c)}(y)),\e
where $S^{(W)}_{E}$ is the classical wormhole action and ${\cal{O}}^{i}(g^{(c)},\phi^{(c)})$ represents a general local action. By summing over the all types of wormhole configurations (see Fig.\ref{mul2}), we obtain the multiverse path integral for the common fields:
\begin{figure}
\begin{center}
\includegraphics[width=8cm]{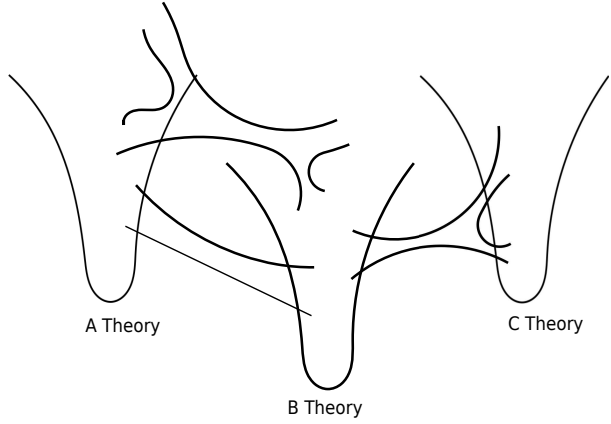}
\caption{An example of typical wormhole configurations. In general, they can have many legs.}
\label{mul2}
\end{center}
\end{figure}
\begin{align}Z&=\sum_{n=0}^{\infty}\int_{{\cal{M}}_{n}} {\cal{D}}g^{(c)}{\cal{D}}\phi^{(c)} \sum_{N_{2}=0}^{\infty}\frac{1}{N_{2}!}\sum_{N_{3}=0}^{\infty}\frac{1}{N_{3}!}\cdots\times\exp\{-S_{E}^{G}(g^{(c)},\phi^{(c)})-S_{E}^{M}(g^{(c)},\phi^{(c)})\}\nonumber\\
&\times[\sum_{i,j}c_{ij}\int dx^{4}\int dy^{4}\sqrt{g(x)}\sqrt{g(y)}{\cal{O}}^{i}(g^{(c)},\phi^{(c)}){\cal{O}}^{j}(g^{(c)},\phi^{(c)})\exp(-2S_{W})]^{N_{2}}\nonumber \\
&\times[\sum_{ijk}c_{ijk}\int d^{4}x\int d^{4}y\int d^{4}z\sqrt{g(x)}\sqrt{g(y)}\sqrt{g(z)}{\cal{O}}^{i}(g^{(c)},\phi^{(c)}){\cal{O}}^{j}(g^{(c)},\phi^{(c)}){\cal{O}}^{k}(g^{(c)},\phi^{(c)})\exp(-3S_{W})]^{N_{3}}\cdot\cdots\nonumber \\
&:=\sum_{n=0}^{\infty}\int_{{\cal{M}}_{n}} {\cal{D}}g{\cal{D}}\phi \sum_{N_{2}=0}^{\infty}\frac{1}{N_{2}!}\sum_{N_{3}=0}^{\infty}\frac{1}{N_{3}!}\cdots\exp\{-S_{E}^{G}(g^{(c)},\phi^{(c)})-S_{E}^{M}(g^{(c)},\phi^{(c)})\}\times[\prod_{i=2}^{\infty}\tilde{S}_{i}^{N_{i}}]\nonumber\\
&=\sum_{n=0}^{\infty}\int_{{\cal{M}}_{n}} {\cal{D}}g^{(c)}{\cal{D}}\phi^{(c)}\exp\{-S_{E}^{G}(g^{(c)},\phi^{(c)})-S_{E}^{M}(g^{(c)},\phi^{(c)})-\sum_{i=2}^{\infty}\tilde{S}_{i}\},\end{align}
where
\be \tilde{S}_{k}:=\sum_{i_{1},i_{2}\cdots,i_{k}}c_{i_{1}i_{2}\cdots i_{k}}\prod_{j=1}^{k}\left(\int d^{4}x_{j}\sqrt{g^{(c)}(x_{j})}{\cal{O}}^{i_{j}}(g^{(c)},\phi^{(c)})\exp(-S_{W})\right):=\sum_{i_{1},i_{2}\cdots,i_{k}}c_{i_{1}i_{2}\cdots i_{k}}S^{(c)}_{i_{1}}S^{(c)}_{i_{2}}\cdots S^{(c)}_{i_{k}},\e
and
\be S^{(c)}_{i}:=\int dx^{4}\sqrt{g^{(c)}(x)}{\cal{O}}^{i}(g^{(c)},\phi^{(c)})\exp(-S_{W}).\label{eq:ordi}\e
Eq.(\ref{eq:ordi}) is an ordinary action. Note that one of $S^{(c)}_{i}$ is the CC produced by the common fields:
\be \int d^{4}x \sqrt{g^{(c)}(x)}\Lambda_{c}.\e
Besides the above common fields, there are fields $\{\phi_{N}\}$ that exist only on N-th universe. We denote their general action by
\be S_{N}(g^{(c)},\phi^{(c)},\phi_{N})=\sum_{j}b_{j}S^{(N)}_{j}(g^{(c)},\phi^{(c)},\phi_{N}).\e
Here $\{b_{j}\}$ are coupling constants, and we here allow interactions between $\phi^{(c)}$ and $\phi_{N}$ in general. Thus, the path integral of the Euclidean multiverse is given by
\be Z^{(E)}_{\text{Multiverse}}=\sum_{n=0}^{\infty}\int_{{\cal{M}}_{n}} {\cal{D}}g^{(c)}{\cal{D}}\phi^{(c)}\left(\prod_{N=1}^{n}{\cal{D}}\phi_{N}\right)\exp\left(-S^{(c)}(g^{(c)},\phi^{(c)})-\sum_{i=2}^{\infty}\tilde{S}_{i}-\sum_{N=1}^{n}S_{N}(g^{(c)},\phi^{(c)},\phi_{N})\right).\label{eq:euclidean}\e
Let us assume that our world is described by the Lorentzian counterpart of Eq.(\ref{eq:euclidean}):
\be Z_{\text{Multiverse}}=\sum_{n=0}^{\infty}\int_{{\cal{M}}_{n}} {\cal{D}}g^{(c)}{\cal{D}}\phi^{(c)}\left(\prod_{N=1}^{n}{\cal{D}}\phi_{N}\right)\exp\left(iS^{(c)}(g^{(c)},\phi^{(c)})+i\sum_{i=2}^{\infty}\tilde{S}_{i}+i\sum_{N=1}^{n}S_{N}(g^{(c)},\phi^{(c)},\phi_{N})\right).\label{eq:Lorentzian}\e
Because $\tilde{S}_{k}$ is the multi-local action, it seems difficult to analyze Eq.(\ref{eq:Lorentzian}). However, if we make the Fourier transform to $\exp\left(i\sum_{i=2}^{\infty}\tilde{S}_{i}\right)$ as a function of $\{S^{(c)}_{k}\}$, we can rewrite Eq.(\ref{eq:Lorentzian}) like the ordinary path integral form:
\begin{align} Z_{\text{Multiverse}}=\sum_{n=0}^{\infty}\int_{{\cal{M}}_{n}} {\cal{D}}g^{(c)}{\cal{D}}\phi^{(c)}&\left(\prod_{N=1}^{n}{\cal{D}}\phi_{N}\right)\int d\overrightarrow{\lambda}\omega(\overrightarrow{\lambda})\nonumber\\
&\times\exp\left(iS^{(c)}(g^{(c)},\phi^{(c)})+i\sum_{N=1}^{n}S_{N}(g^{(c)},\phi^{(c)},\phi_{N})+i\sum_{k}\lambda_{k}S^{(c)}_{k}\right)\nonumber\end{align}
\begin{align}\h{1.8cm}&=\sum_{n=0}^{\infty}\int_{{\cal{M}}_{n}} {\cal{D}}g^{(c)}{\cal{D}}\phi^{(c)}\left(\prod_{N=1}^{n}{\cal{D}}\phi_{N}\right)\int d\overrightarrow{\lambda}\omega(\overrightarrow{\lambda})e^{iS(g^{(c)},\phi^{(c)},\{\phi_{N}\})}\nonumber\\
&=\int d\overrightarrow{\lambda}\omega(\overrightarrow{\lambda})\sum_{n=0}^{\infty}\left(\prod_{i=1}^{n}Z^{(i)}_{\text{universe}}\right), \end{align}
where
\be S(g^{(c)},\phi^{(c)},\{\phi_{N}\})=S^{(c)}(g^{(c)},\phi^{(c)})+\sum_{N=1}^{n}S_{N}(g^{(c)},\phi^{(c)},\phi_{N})+\sum_{k}\lambda_{k}S^{(c)}_{k}\e
is the total effective action, $\omega(\overrightarrow{\lambda})$ is the Fourier coefficient, and 
\be Z_{\text{universe}}^{(i)}=\int_{\text{universe}}{\cal{D}}g^{(c)}{\cal{D}}\phi^{(c)}{\cal{D}}\phi_{i}\exp\left(iS^{(c)}(g^{(c)},\phi^{(c)})|_{\text{universe}}+i\sum_{k}\lambda_{k}S^{(c)}_{i}|_{\text{universe}}+iS_{i}(g^{(c)},\phi^{(c)},\phi_{i})\right)\label{eq:single path}\e
is the path integral of the $i$-th universe. Note that the coupling constants of a single universe are given by $\{b_{i},\lambda_{i}\}$. Especially, the effective CC of the N-th universe is
\be \bar{\Lambda}_{N}=\Lambda_{N}+\Lambda_{c},\label{eq:effective cosmological constant}\e
where we have denoted the effective cosmological constant produced by the fields $\{\phi_{N}\}$ as $\Lambda_{N}$. In the following discussion, we assume that the universe can be described by its scale factor $a$ and its effective potential $V(a)$. In this case, Eq.(\ref{eq:single path}) becomes
\be Z^{(i)}_{universe}(a_{f},a_{i})=\int{\cal{D}}p_{a}\int_{t=0 ,a(0)=a_{i}}^{t=1 ,a(1)=a_{f}}{\cal{D}}a{\cal{D}}N \exp\left(i\int_{0}^{1} dt(p_{a}\dot{a}-N{\cal{H}}^{(i)}(\lambda))\right),\label{eq:scale path}\e
where $N(t)$ is the lapse function, and ${\cal{H}}^{(i)}(\lambda)=-\frac{G}{a}\{\frac{1}{2}p^{2}_{a}+V^{(i)}(a)\}$ is the Hamiltonian of the i-th universe. By choosing the gauge such that $N(t)$ is a constant and make the variable change $t\rightarrow Tt$, Eq.(\ref{eq:scale path}) becomes
\begin{align} Z^{(i)}_{universe}(a_{1},a_{0})&=\int_{-\infty}^{\infty} dT\int_{t=0 ,a(0)=a_{0}}^{t=T ,a(T)=a_{1}}{\cal{D}}a \exp\left(i\int_{0}^{T} dt(p_{a}\dot{a}-N{\cal{H}}^{(i)}(\lambda))\right)\nonumber\\
&=\int_{-\infty}^{\infty}dT\langle a_{1}|\exp(-i{\cal{H}}^{(i)}(\lambda)T)|a_{0}\rangle\nonumber\\
&=\langle a_{1}|\delta({\cal{H}}^{(i)}(\lambda))|a_{0}\rangle\nonumber\\
&=\langle a_{1}|\delta({\cal{H}}^{(i)}(\lambda))\int_{-\infty}^{\infty}|E\rangle\langle E||a_{0}\rangle\nonumber\\
&=\langle a_{1}|E=0\rangle\langle E=0|a_{0}\rangle:=\phi^{(i)}_{E=0}(a_{1})\phi^{*(i)}_{E=0}(a_{0}),\label{eq:scale path2}\end{align}
where we have inserted the complete set
\be 1=\int_{-\infty}^{\infty}|E\rangle \langle E|,\e
\be {\cal{H}}^{(i)}(\lambda)|E\rangle=E|E\rangle.\e
The WKB solution of the zero eigenfunction $\phi^{(i)}_{E=0}(a)$ for large $a$ is given by  \cite{Kawai:2011rj,Kawai:2011qb,Kawai:2013wwa,Hamada:2014ofa}
\be \phi^{(i)}_{E=0}(a)\simeq\frac{1}{(\bar{\Lambda}_{i})^{1/4}a^{1/2}}\exp(i\cdots),\h{5mm}(\text{for $\bar{\Lambda}_{i}>0$})\label{eq:WKB}\e
where $\bar{\Lambda}_{i}$ is the total cosmological constant. Note that Eq.(\ref{eq:WKB}) becomes large when $\bar{\Lambda}_{i}$ is close to zero. 

\section{Wave Function of Multiverse}
In this section, we define the wave function of the multiverse. In general, there can be many effective theories in which a single universe can live. We label these different theories by $\{\alpha\}$ for simplicity. Furthermore, there can be many vacuums within a theory, and we distinguish them by $\{j\}$.  Although we do not well understand how a single universe emerges from nothing, we assume that this can be represented effectively by the probability amplitude $p_{\alpha}(j,K)$. Namely, $p_{\alpha}(j,K)$ is the amplitude with which a universe having a theory $\alpha$, vacua j and topology K emerges. Of course, $p_{\alpha}(k,K)$ should satisfy
\be \sum_{\alpha,k,K}|p_{\alpha}(k,K)|^{2}=1.\e
Using $p_{\alpha}(k,K)$, we can define the quantum state of the n universes as follows:
\begin{align} |\Psi_{n}\rangle=\int d\overrightarrow{\lambda}p_{\alpha_{1}}(k_{1},K_{1})p_{\alpha_{2}}(k_{2},K_{2})\cdots p_{\alpha_{n}}&(k_{n},K_{n})|\epsilon_{1}\rangle|\text{vacuum} \h{1mm}k_{1}\rangle\otimes|\epsilon_{2}\rangle|\text{vacuum}\h{1mm}k_{2}\rangle\otimes\nonumber\\
&\cdots\otimes|\epsilon_{n}\rangle|\text{vacuum}\h{1mm} k_{n}\rangle\otimes\omega(\overrightarrow{\lambda})|\overrightarrow{\lambda}\rangle,\end{align}
where we have assumed that the initial state of a universe is given by the eigenstate of the scale factor $|\epsilon\rangle$. Note that we should take into account the Hilbert space of $\{\lambda_{i}\}$ because they are variables now. By using the path integral (\ref{eq:scale path2}), we can obtain the wave function of the n universes having coupling constants $\lambda_{i}$ as follows:
\begin{align} \Psi_{n}(\overrightarrow{\lambda},a_{1},\cdots,a_{n})&=\langle a_{1}|\cdots\langle a_{n}|\langle\overrightarrow{\lambda}\|\Psi_{n}\rangle\nonumber\\
&=\omega(\overrightarrow{\lambda})\prod_{i=1}^{n}p_{\alpha_{i}}(k_{i},K_{i})Z^{(i)}_{\text{universe}}(\overrightarrow{\lambda},a_{i},\epsilon_{i})\nonumber\\
&=\omega(\overrightarrow{\lambda})\prod_{i=1}^{n}p_{\alpha_{i}}(k_{i},K_{i})\times\phi_{E=0}^{(i)}(a_{i})\phi_{E=0}^{*(i)}(\epsilon_{i}).\end{align}
We can obtain the probability distribution $P$ of coupling constants by tracing out the number of other universes and their scale factors:
\begin{align} P(\overrightarrow{\lambda},\alpha_{0},k_{0},K_{0}):&=\sum_{n=1}^{\infty}\frac{1}{n!}\sum_{K_{1},\cdots,K_{n}}\sum_{\alpha_{1}\cdots \alpha_{n}}\sum_{k_{1}\cdots k_{n}}\int\cdots\int \left(\prod_{i=0}^{n}da_{i}\right)\Psi^{*}_{n+1}(\lambda,a_{0},\cdots,a_{n})\Psi_{n+1}(\lambda,a_{0},\cdots,a_{n})\nonumber\\
&=|p_{\alpha_{0}}(k_{0},K_{0})\omega(\overrightarrow{\lambda})|^{2}|\phi^{(\alpha_{0},k_{0},K_{0})}_{E=0}(\epsilon_{0})|^{2}\times\int da|\phi^{(\alpha_{0},k_{0},K_{0})}_{E=0}(a)|^{2}\nonumber\end{align}
\be \h{3cm}\times\sum_{n=1}^{\infty}\frac{1}{n!}\left(\sum_{K}\sum_{\alpha}\sum_{k}|p_{\alpha}(k,K)\phi^{(\alpha,k,K)}_{E=0}(\epsilon)|^{2}\times\int da|\phi^{(\alpha,k,K)}_{E=0}(a)|^{2}\right)^{n}\nonumber\e
\begin{align}\h{2cm}=|p_{\alpha_{0}}(k_{0},K_{0})\omega(\overrightarrow{\lambda})|^{2}&\times|\phi^{(\alpha_{0},k_{0},K_{0})}_{E=0}(\epsilon_{0})|^{2}\times\int da|\phi^{(\alpha_{0},k_{0},K_{0})}_{E=0}(a)|^{2}\nonumber\\
&\times\exp\left(\sum_{K}\sum_{\alpha}\sum_{k}|\phi^{(\alpha,k,K)}_{E=0}(\epsilon)p_{\alpha}(k,K)|^{2}\times\int da|\phi^{(\alpha,k,K)}_{E=0}(a)|^{2}\right)\nonumber\\
&:=f(\overrightarrow{\lambda})\times\exp(F(\overrightarrow{\lambda})),\label{eq:coupling pro}\end{align}
where
\be f(\overrightarrow{\lambda}):=|p_{\alpha_{0}}(k_{0},K_{0})\omega(\overrightarrow{\lambda})|^{2}\times|\phi^{(\alpha_{0},k_{0},K_{0})}_{E=0}(\epsilon_{0})|^{2}\times\int da|\phi^{(\alpha_{0},k_{0},K_{0})}_{E=0}(a)|^{2},\e
and
\be F(\overrightarrow{\lambda}):=\sum_{K}\sum_{\alpha}\sum_{k}|\phi^{(\alpha,k,K)}_{E=0}(\epsilon)p_{\alpha}(k,K)|^{2}\int da|\phi^{(\alpha,k,K)}_{E=0}(a)|^{2}.\e
Here, $0$ index represents our universe. Eq.(\ref{eq:coupling pro}) has strong peaks where $F(\overrightarrow{\lambda})$ becomes large. As well as the previous works \cite{Kawai:2011rj,Kawai:2011qb,Kawai:2013wwa,Hamada:2014ofa,Hamada:2014xra}, we understand that the coupling constants of our universe are fixed at the point in the $\{\lambda_{i}\}$ space where Eq.(\ref{eq:coupling pro}) becomes maximum\footnote{This is very similar to the relation between a system and its reserver in statistical mechanics; the temperature of a system is determined by the interaction with the resorver. See \ref{app:stati}.}. Generally, there are many local maximums of $F(\overrightarrow{\lambda})$ because each universe has its own peak in $F(\overrightarrow{\lambda})$. These peaks may correspond to our universe's peak such that the cosmological constant of our universe becomes small. Although it is very difficult to analyze Eq.(\ref{eq:coupling pro}) generally, we can consider a few simple cases.\\
\\
Case(I-A): \underline{There is only an unique effective theory, and this theory has an unique vacuum.}\\
This is the simple case discussed in \cite{Kawai:2011rj,Kawai:2011qb,Kawai:2013wwa,Hamada:2014ofa,Hamada:2014xra}, and the probability distribution becomes
\be |\omega(\overrightarrow{\lambda})\phi^{(K_{0})}_{E=0}(\epsilon_{0})|^{2}\times\int da|\phi^{(K_{0})}_{E=0}(a)|^{2}\times\exp\left(\sum_{K}|p(K)\phi^{(K)}_{E=0}(\epsilon)|^{2}\times\int da|\phi^{(K)}_{E=0}(a)|^{2}\right).\label{eq:unique1}\e
Eq.(\ref{eq:unique1}) has a maximum such that the integral 
\be \int da|\phi^{(K)}_{E=0}(a)|^{2}\label{eq:integral}\e
becomes maximum. Actually, since the WKB solution of the wave function of the universe is given by Eq.(\ref{eq:WKB}), one can see that Eq.(\ref{eq:integral}) has a maximum where the cosmological constant becomes zero. This is the solution to the CCP suggested in \cite{Kawai:2011rj,Kawai:2011qb,Kawai:2013wwa,Hamada:2014ofa,Hamada:2014xra}. 
\\
\\
Case(I-B): \underline{There is only an unique effective theory, and this theory has many vacuums.}\\
In this case, $F(\overrightarrow{\lambda})$ becomes
\be F(\overrightarrow{\lambda})=\sum_{K}\sum_{k}|\phi^{(k,K)}_{E=0}(\epsilon)p(k,K)|^{2}\int da|\phi^{(k,K)}_{E=0}(a)|^{2}.\label{eq:unique2}\e
\newpage
\begin{figure}[!h]
\begin{center}
\includegraphics[width=7cm]{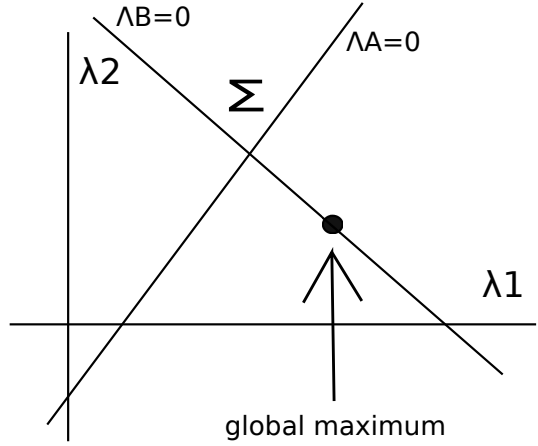}
\caption{An example of the hypersurfaces of Case(II). The global maximum is not necessary on the hypersurface such that the cosmological constant of our universe becomes zero.}
\label{lambda1}
\end{center}
\end{figure}
\noindent The cosmological constant of a single universe living in the vacuum k is given by Eq.(\ref{eq:effective cosmological constant});
\be \bar{\Lambda}_{k}=\Lambda_{k}+\Lambda_{c}.\nonumber\e
Using the WKB solution (\ref{eq:WKB}), one can see that the integral
\be \int da|\phi^{(k,K)}_{E=0}(a)|^{2}\label{eq:integral2}\e
has a peak where  $\bar{\Lambda}_{k}\rightarrow0$. Thus, if $p(k,K)$ dose not so change between universes having different vacuums, each universe has a similar maximum in $F(\overrightarrow{\lambda})$ because they live in the same effective theory \footnote{Of course, details are different between universes. }. 
However, if $p(k,K)$ changes drastically between universes, this mechanism fails to succeed in general.
\\
\\
Case(II): \underline{There are only two effective theories, and each theory has an unique vacuum.}\\
We denote these two theories by A and B, and assume that we live in the A theory (the SM). In this case, $F(\{\lambda_{i}\})$ becomes
\begin{align} F(\overrightarrow{\lambda})&=\sum_{K}\sum_{\alpha=A,B}|\phi^{(\alpha,K)}_{E=0}(\epsilon)p_{\alpha}(K)|^{2}\int da|\phi^{(\alpha,K)}_{E=0}(a)|^{2}\nonumber\\
&=\sum_{K}\left(|p_{A}(K)\phi^{(A,K)}_{E=0}(\epsilon)|^{2}\int da|\phi^{(A,K)}_{E=0}(a)|^{2}+|p_{B}(K)\phi^{(B,K)}_{E=0}(\epsilon)|^{2}\int da|\phi^{(B,K)}_{E=0}(a)|^{2}\right).\label{eq:case2}\end{align}
For simplicity, let us assume that $p_{i}(K)=$constant. As well as the Case(I), $F(\overrightarrow{\lambda})$ has maximums where $\bar{\Lambda}_{A}$ or $\bar{\Lambda}_{B}$ become zero. They define two hypersurfaces $\Sigma_{A}$ and $\Sigma_{B}$ in the $\{\lambda_{i}\}$ space. If a global maximum exists on the intersecting surface $\Sigma=\Sigma_{A}\cap\Sigma_{B}$, we can conclude that $\bar{\Lambda}_{A}=\Lambda_{SM}$ becomes zero. However, if  the global maximum exists on $\Sigma_{B}\setminus\Sigma$, $\bar{\Lambda}_{A}$ does not necessary become small. 
Namely, if the maximum of the other universe is stronger than that of our universe, $\{\lambda_{i}\}$ are fixed so that the integral of the other universe 
\be \int da|\phi^{(B,K)}_{E=0}(a)|^{2}\e
becomes maximum. As a result, we can not claim the Big Fix of our universe in this case. One can easily understand that this conclusion is rather general. To make more quantitative argument, we need to know the other universe's history, but this is a difficult question.\\
\section{Summary}
In this paper, we have discussed the quantum mechanics of multiverse. We have defined the path integral and the wave function of multiverse where each universe can have a different effective theory and vacuum. From this point of view, the previous and recent works \cite{Kawai:2011rj,Kawai:2011qb,Kawai:2013wwa,Hamada:2014ofa,Hamada:2014xra} can be understood as the simple case such that there is a unique effective theory and vacuum. We have introduced the probability amplitude $p_{\alpha}(k,K)$ and assumed that it does not depend on parameters $\{\lambda_{i}\}$ because $p_{\alpha}(k,K)$ should be determined by a microscopic theory.  We have also considered a few special cases, and showed that $\Lambda_{SM}$ is fixed very  small for the Case(I-A) (and (I-B)), but in more general cases, Eq.(\ref{eq:coupling pro}) does not necessary have a maximum where $\Lambda_{SM}$ becomes small. It is very difficult to analyze whether Eq.(\ref{eq:coupling pro}) actually has such a maximum because we must know the histories of other universes. We think that it might be interesting to consider Case(II) in detail by assuming that the other theory (B) is similar to the SM. At any rate, the theory of multiverse and the wormholes is one of the interesting possibilities to solve the fine tuning problems.

\section*{Acknowledgement} 
We thank Hikaru Kawai and Yuta Hamada for valuable discussions.
\appendix
\def\thesection{Appendix \Alph{section}}
\section{Analogy with Statistical Mechanics}\label{app:stati}
In the statistical  mechanics, the thermal equilibrium of the system having the temperature $T$ can be realized by the interaction between the system and the reservoir. Furthermore, $T$ is determined by the property of the reservoir. Namely,  if we denote the number of state of  the reservoir having the energy $E$ by
\be \Omega_{R}(E),\e
the temperature of the system is given by
\be T^{-1}=\frac{\partial}{\partial E}\log(\Omega_{R}(E))|_{\bar{E}}\e
where $\bar{E}$ is the total energy of the system and the reservoir. One can see that this is very similar to the above argument of multiverse.  In fact, for the case of multiverse, the other universes play roles as a reservoir, and the coupling constants of our universe are determined by solving the equation
\be \frac{\partial P(\overrightarrow{\lambda})}{\partial \lambda_{i}}=0,\e
or, by using Eq.(\ref{eq:coupling pro}) 
\be \frac{\partial f(\overrightarrow{\lambda})}{\partial \lambda_{i}}+f(\overrightarrow{\lambda})\frac{\partial F(\overrightarrow{\lambda})}{\partial \lambda_{i}}=0.\label{eq:diff}\e 
Thus, we can interpret that we live in the thermal bath of multiverse, and the coupling constants are like the temperature. Here, the interactions between other universes are caused by wormholes. For the Case(I-A), because
\be \frac{f(\overrightarrow{\lambda})}{|\omega(\overrightarrow{\lambda})|^{2}}=F(\overrightarrow{\lambda}),\e
Eq.(\ref{eq:diff}) leads to
\be |\omega(\overrightarrow{\lambda})|^{2}\left(1+F(\overrightarrow{\lambda})\right)\frac{\partial F(\overrightarrow{\lambda})}{\partial \lambda_{i}}+F(\overrightarrow{\lambda})\frac{\partial |\omega(\overrightarrow{\lambda})|^{2}}{\partial \lambda_{i}}=0.\e
If we can neglect the second term, the coupling constants of our universe is simply determined by
\be \frac{\partial F(\overrightarrow{\lambda})}{\partial \lambda_{i}}=0\h{4mm}(\text{for the Case(I-A)}).\e



\begin{thebibliography}{4}
\bibitem{Coleman:1988tj} 
  S.~R.~Coleman,
  ``Why There Is Nothing Rather Than Something: A Theory of the Cosmological Constant,''
  Nucl.\ Phys.\ B {\bf 310}, 643 (1988).


\bibitem{Kawai:2011rj}
  H.~Kawai and T.~Okada,
  {\it``Asymptotically Vanishing Cosmological Constant in the Multiverse,''}
  Int.\ J.\ Mod.\ Phys.\ A {\bf 26} (2011) 3107
  [arXiv:1104.1764 [hep-th]].


\bibitem{Kawai:2011qb} 
  H.~Kawai and T.~Okada,
  ``Solving the Naturalness Problem by Baby Universes in the Lorentzian Multiverse,''
  Prog.\ Theor.\ Phys.\  {\bf 127}, 689 (2012)
  [arXiv:1110.2303 [hep-th]].
  
\bibitem{Kawai:2013wwa} 
  H.~Kawai,
  ``Low energy effective action of quantum gravity and the naturalness problem,''
  Int.\ J.\ Mod.\ Phys.\ A {\bf 28}, 1340001 (2013).
  
\bibitem{Hamada:2014ofa} 
  Y.~Hamada, H.~Kawai and K.~Kawana,
  ``Evidence of the Big Fix,''
  arXiv:1405.1310 [hep-ph].

\bibitem{Hamada:2014xra} 
  Y.~Hamada, H.~Kawai and K.~Kawana,
  ``Weak Scale From the Maximum Entropy Principle,''
  PTEP {\bf 2015}, no. 3, 033B06 (2015)
  [arXiv:1409.6508 [hep-ph]].


\bibitem{Chatrchyan:2012ufa}
  S.~Chatrchyan {\it et al.}  [CMS Collaboration],
  {\it``Observation of a new boson at a mass of 125 GeV with the CMS experiment at the LHC,''}
  Phys.\ Lett.\ B {\bf 716} (2012) 30
  [arXiv:1207.7235 [hep-ex]].

\bibitem{Aad:2012tfa}
  G.~Aad {\it et al.}  [ATLAS Collaboration],
  {\it``Observation of a new particle in the search for the Standard Model Higgs boson with the ATLAS detector at the LHC,''}
  Phys.\ Lett.\ B {\bf 716} (2012) 1
  [arXiv:1207.7214 [hep-ex]].
  
\bibitem{Hamada:2012bp} 
  Y.~Hamada, H.~Kawai and K.~-y.~Oda,
  ``Bare Higgs mass at Planck scale,''
  Phys.\ Rev.\ D {\bf 87}, no. 5, 053009 (2013)
  [arXiv:1210.2538 [hep-ph]].

\bibitem{Hamada:2013cta} 
  Y.~Hamada, H.~Kawai and K.~-y.~Oda,
  ``Bare Higgs mass and potential at ultraviolet cutoff,''
  arXiv:1305.7055 [hep-ph].

\bibitem{Hamada:2013mya} 
  Y.~Hamada, H.~Kawai and K.~-y.~Oda,
  ``Minimal Higgs inflation,''
  PTEP {\bf 2014}, 023B02 (2014)
  [arXiv:1308.6651 [hep-ph]].


\bibitem{Hamada:2014iga} 
  Y.~Hamada, H.~Kawai, K.~-y.~Oda and S.~C.~Park,
  ``Higgs inflation still alive,''
  arXiv:1403.5043 [hep-ph].

\bibitem{Hamada:2015ria} 
  Y.~Hamada, H.~Kawai and K.~y.~Oda,
  ``Eternal Higgs inflation and cosmological constant problem,''
  arXiv:1501.04455 [hep-ph].
  
\bibitem{Froggatt:1995rt} 
  C.~D.~Froggatt and H.~B.~Nielsen,
  ``Standard model criticality prediction: Top mass 173 +- 5-GeV and Higgs mass 135 +- 9-GeV,''
  Phys.\ Lett.\ B {\bf 368}, 96 (1996)
  [hep-ph/9511371].

  C.~D.~Froggatt, H.~B.~Nielsen and Y.~Takanishi,
  ``Standard model Higgs boson mass from borderline metastability of the vacuum,''
  Phys.\ Rev.\ D {\bf 64}, 113014 (2001)
  [hep-ph/0104161];

  H.~B.~Nielsen,
  ``PREdicted the Higgs Mass,''
  arXiv:1212.5716 [hep-ph];
  
  Y.~Hamada, H.~Kawai and K.~y.~Oda,
  ``Predictions on mass of Higgs portal scalar dark matter from Higgs inflation and flat potential,''
  JHEP {\bf 1407}, 026 (2014)
  [arXiv:1404.6141 [hep-ph]];
  
  K.~Kawana,
  ``Multiple Point Principle of the Standard Model with Scalar Singlet Dark Matter and Right Handed Neutrinos,''
  PTEP {\bf 2015}, no. 2, 023B04 (2015)
  [arXiv:1411.2097 [hep-ph]].
  
\bibitem{Meissner:2006zh} 
  K.~A.~Meissner and H.~Nicolai,
  ``Conformal Symmetry and the Standard Model,''
  Phys.\ Lett.\ B {\bf 648}, 312 (2007)
  [hep-th/0612165];
  K.~A.~Meissner and H.~Nicolai,
  ``Effective action, conformal anomaly and the issue of quadratic divergences,''
  Phys.\ Lett.\ B {\bf 660}, 260 (2008)
  [arXiv:0710.2840 [hep-th]];
  S.~Iso, N.~Okada and Y.~Orikasa,
  ``Classically conformal $B^-$ L extended Standard Model,''
  Phys.\ Lett.\ B {\bf 676}, 81 (2009)
  [arXiv:0902.4050 [hep-ph]];
  S.~Iso, N.~Okada and Y.~Orikasa,
  ``The minimal B-L model naturally realized at TeV scale,''
  Phys.\ Rev.\ D {\bf 80}, 115007 (2009)
  [arXiv:0909.0128 [hep-ph]];
  H.~Aoki and S.~Iso,
  ``Revisiting the Naturalness Problem -- Who is afraid of quadratic divergences? --,''
  Phys.\ Rev.\ D {\bf 86}, 013001 (2012)
  [arXiv:1201.0857 [hep-ph]];
  S.~Iso and Y.~Orikasa,
  ``TeV Scale B-L model with a flat Higgs potential at the Planck scale - in view of the hierarchy problem -,''
  PTEP {\bf 2013}, 023B08 (2013)
  [arXiv:1210.2848 [hep-ph]];
  M.~Hashimoto, S.~Iso and Y.~Orikasa,
  ``Radiative symmetry breaking at the Fermi scale and flat potential at the Planck scale,''
  Phys.\ Rev.\ D {\bf 89}, 016019 (2014)
  [arXiv:1310.4304 [hep-ph]];
  M.~Hashimoto, S.~Iso and Y.~Orikasa,
  ``Radiative Symmetry Breaking from Flat Potential in various U(1)' models,''
  Phys.\ Rev.\ D {\bf 89}, 056010 (2014)
  [arXiv:1401.5944 [hep-ph]];
  R.~Foot, A.~Kobakhidze, K.~L.~McDonald and R.~R.~Volkas,
  ``A Solution to the hierarchy problem from an almost decoupled hidden sector within a classically scale invariant theory,''
  Phys.\ Rev.\ D {\bf 77}, 035006 (2008)
  [arXiv:0709.2750 [hep-ph]].

\bibitem{Kawana:2015tka} 
  K.~Kawana,
  ``Criticality and Inflation of the Gauged B-L Model,''
  arXiv:1501.04482 [hep-ph].

\bibitem{Shaposhnikov:2009pv} 
  M.~Shaposhnikov and C.~Wetterich,
  ``Asymptotic safety of gravity and the Higgs boson mass,''
  Phys.\ Lett.\ B {\bf 683}, 196 (2010)
  [arXiv:0912.0208 [hep-th]].

\bibitem{Garriga:2012bc} 
  J.~Garriga and A.~Vilenkin,
  ``Watchers of the multiverse,''
  JCAP {\bf 1305}, 037 (2013)
  [arXiv:1210.7540 [hep-th]].

\bibitem{Vilenkin:2013ik} 
  A.~Vilenkin,
  ``Global structure of the multiverse and the measure problem,''
  AIP Conf.\ Proc.\  {\bf 1514}, 7 (2012)
  [arXiv:1301.0121 [hep-th]].

\bibitem{Vilenkin:2013loa} 
  A.~Vilenkin,
  ``A quantum measure of the multiverse,''
  arXiv:1312.0682 [hep-th].



\end{thebibliography}
\end{document}